\documentstyle[12pt,psfig]{article} 
\font\elevenbf=cmbx10 scaled\magstep 1                                        
\begin{document}                                                              
%\begin{flushright}{LNF-94/xxx P}  
%\end{flushright}                                                      
\begin{center}                                                                
{ \large\bf Collision times in $\pi \pi$ and $\pi K$ scattering and 
spectroscopy of meson resonances\\}
\vskip 2cm  
{ N. G. Kelkar$^1$, M. Nowakowski$^1$ and K. P. Khemchandani$^2$\\} 
$^1$Departamento de Fisica, Universidad de los Andes,
Cra.1 No.18A-10, Santafe de Bogota, Colombia\\
$^2$Nuclear Physics Division, Bhabha Atomic Research Centre, Mumbai 400085
\end{center}
\vskip .5cm
\begin{center}
\end{center}                             
\begin{abstract}
Using the concept of collision time (time delay) introduced by 
Eisenbud and Wigner and its connection to on-shell intermediate 
unstable states, we 
study mesonic resonances in $\pi \pi$ and $\pi K$ scattering. The time-delay 
method proves its usefulness by revealing the spectrum of the well-known $\rho$
- and $K^*$- mesons and by supporting some speculations on $\rho$-mesons 
in the $1200$ MeV region. We use this method further to shed some light on more
speculative meson resonances, among others the enigmatic scalars. We confirm 
the existence of chiralons below $1$ GeV in the unflavoured and strange meson
sector.
\end{abstract}                                                                
\newpage 
\section{Introduction}
The study of hadronic resonances, be it baryons or mesons, has never ceased 
(with seasonal fluctuations) to interest the particle 
and nuclear physics community for several specific and one global reason 
\cite{review,oset1,montanet,klempt}. 
The latter reason is to be found in the fact
that hadrons and their interaction are the low energy manifestations of QCD.
The specific reasons include among others: 
(i) the wish
to improve upon standard results (e.g. resonance parameter extraction), (ii)
the attempt to understand inconsistencies and puzzles, if any, (iii) possible 
new resonances (of speculative status, enigmatic nature, 
predicted by models and/or resurrection of previously refuted results), 
(iv) exotic resonances 
like glueballs, hybrids (named also hermaphrodites and meiktons), 
multiquark states such as $q^2\bar{q}^2$, mesonic molecules 
and $J^{PC}$-exotics, 
(v) the enigmatic scalar sector with the return of the $\sigma$-meson 
into the data book of the Particle Data Group (PDG)
\cite{pdg1}. We shall come back to (i)-(v) in the next section.
Here it suffices to note that it is very often the existence of a resonance 
which is at stake. Insufficient data, weak signal and different methods 
applied to analyze data can result in different conclusions. 
The analytical methods
to extract resonances and their parameters are well known and well 
documented. It then comes as a mild surprise to know that not {\it all} 
known analytical 
methods have been fully exploited. For instance, it is now fifty years since
the concept of
collision time (time delay) was introduced by Eisenbud and Wigner
\cite{wigner,wigner2,wigner3}. Its 
connection to resonances is best explained in Wigner's own words
\cite{wigner} : 
`` Close to resonances, where the incident particle is in fact captured and 
retained for some time by the scattering centre, 
${d \eta \over d k}$ [this is the collision time where $\eta$ is
the scattering phase shift] will assume large positive 
values''. Not much has changed since then, 
except that the concept of time delay has been further elaborated by many 
%authors \cite{earlypapers,smith,osborn,ohanian}, 
%incorporated into text books \cite{brans,books,goldberger,baz,joa,taylor} 
authors [9-12], 
incorporated into text books [13-18] 
and newly rediscovered in a different context \cite{newpapers,twonew}. 
In all these references the connection to 
resonance physics is well documented and stressed. Nowadays, one 
might not look upon a resonance as a particle being retained by
another for a while, but rather as an unstable intermediate
state characterized by several quantum numbers. 
To clarify this point, we quote ref. \cite{ohanian}: 
`` Finally we remark that a sharp maximum in the time delay 
(essentially condition (iii) [the specific time delay 
${d \delta \over dk}$ has a sharp maximum])
is sufficient for the existence of a resonance; the argument 
is given in Ref. 12 [our reference \cite{goldberger}]''. 
We could continue quoting
text books and papers on time delay, but it would mean 
repeating more or less one and
the same phrase. The surprise then grows when we learn that only recently 
the time delay concept has been put to a test in connection with hadron 
resonances
\cite{ng,ngandme}, with reasonable success. By this we do not mean that
time delay has not been extensively discussed in literature 
(see earlier references and for some
criticism see \cite{dalitz}), 
but we are not aware of any work which has made a 
practical use of the advocated time delay in connection with hadron
resonances. Time delay analysis for $\Sigma$-hypernuclear states has been
successfully performed in \cite{sigmahyp}. 
A related and less documented but somewhat popular 
concept, namely the {\it speed plot}, is occasionally found in literature
\cite{pdg1,hoehler,yogi}. 
However, {\it speed plots} are positive by definition  
whereas time delay (as obvious from the citation from Wigner's paper) 
can assume positive as well as negative values. Hence, regions 
with a negative bump in the collision time, could   
come out as positive peaks in the speed plot. 
In this context, we quote Wigner once again: 
``One would expect 
(on the basis of Liouville theorem or completeness relation) that the 
two effects [positive and negative values of time delay], 
on the whole, balance each other, i.e. that the integral of 
${d \eta \over dk}$ (in Wigner's notation, $\eta$ is phase
shift, which is denoted as $\delta$ in the present work) over the whole 
energy range is close to zero, .....
Hence, if the cross section shows a resonance behaviour, one will 
expect $\eta$ to decrease slowly between resonances and increase fast at 
resonances, increase and decrease almost exactly balancing 
if considered over the whole energy spectrum.''  
A demonstration of the above statement can actually be found in 
\cite{lifshutz} for a constructed example of seven consecutive 
Breit-Wigner resonances. A detailed comparison between time delay and
speed plots can be found in \cite{we3}.  
We supplement the above by noting that the connection between time delay and 
the statistical density of states in scattering has been given in 
\cite{statistics}. We also note that not every peak in the cross
section can be attributed to a resonance \cite{ohanian}. 

In view of the above discussion, it seems worth undertaking 
an examination of time delay in scattering processes with data, 
encountering several resonances over a wide energy range. 
In \cite{ng, ngandme} we have applied the time delay method 
for baryon resonances in $\pi N$ scattering. In the present paper, 
we do the analog for mesonic resonances in $\pi \pi$ and $\pi K$ collisions. 
Indeed, we get very sharp {\it positive} peaks for all established 
resonances which confirms the statements made by 
Wigner and others in literature. 
Agreement with the known resonances adds support to the interpretation
of the additional peaks we find as new resonances
(indeed, these additional peaks have been a matter of debate for some time).

Our paper is organized as follows: in the next section we give some necessary 
facts about mesonic resonances. These facts will be necessary to interpret our 
results. In the subsequent section we briefly discuss the relevant 
formulae on time delay. In sections 4 and 5 we present the 
collision times calculated for $\pi \pi$ and $\pi K$ scattering. 
The last section is devoted to conclusions.

\section{A short survey of relevant mesonic resonances}
Given the seniority of the subject, the number of possible topics worth 
mentioning, even with the restriction to the points (i)-(v) 
mentioned in the Introduction in context with the continuing interest 
in mesonic resonances, is of course too large to be covered here. 
Therefore the short survey 
below is coloured by what we thought to be relevant for the results of the
present paper.

As an example of the points (i) and (ii) from the previous section, we quote 
the case of the $\rho$ mesons. Till the year 2000, the PDG listed a group of 
$J^{PC}=1^{--}$ mesons with masses around $1100-1200$ MeV discovered in 
$e^+e^-$ collisions \cite{pdg2}. In the 2002 edition \cite{pdg1} this entry 
has been removed due to the lack of further evidence supporting the old data. 
In the same year this controversy has been revived by an experimental 
indication of an isovector state with mass around $1200$ MeV 
\cite{achasov,pick,donnachie1}. As noted in
\cite{donnachie1} this is also supported by the $\gamma p \to (\omega \pi) p$
reaction, where one finds a mass enhancement in the $\omega \pi$ system which 
is partly attributed to additional $\rho$ meson \cite{rhofurther}.

Additional conclusions regarding the $\rho$ mesons are drawn from 
`inconsistencies' in $e^+ e^-$ collisions and $\tau$ decays 
\cite{donnachie2, donnachie3}.
These seem to indicate the necessity of a vector hybrid 
\cite{vechyb,donnachie2} which serves as a good 
example for the points (iii)-(iv) in the $\rho$ mesonic sector.
It is beyond the scope of this work to describe in detail the situation 
of exotic mesons \cite{exotics}. 
However, one of the exotics, namely the glueball, has indirectly to do with 
the findings of our paper. We say so because, according to lattice theory 
and other models, the lightest glueball is predicted to be a true scalar with 
a mass around $1700$ MeV \cite{lattice}.
Opinion as to which one of the low lying scalars has 
the largest admixture of glue differs \cite{glue,ochs,anisovich}.  
This adds to the other problems encountered in the scalar 
sector (point (v)). Regarding the glue, 
in the scalar sector, the interpolating field for $\sigma$
could be ${\bf F} \cdot {\bf F}$ \cite{ellis}.  
The situation is not unlike the $\eta$ mesons, 
where through the $U(1)_A$ anomaly,  
there is a connection between ${\bf F}\cdot {\bf \tilde{F}}$ and   
the $\eta_0$ field \cite{eta} making us think
about a glue content of eta mesons and $\eta_0$-$\eta_8$-glue mixing 
\cite{etaglue}. 
In any case, the most famous problem beside the glue in the scalar 
sector is the $\sigma$ meson 
itself, which was removed from PDG in 1974 and reappeared there much later.
This metamorphosis \cite{svec} (of appearance and disappearance)
is typical for this meson in other respects too. From the 
experimental point of view,
it is sometimes claimed that this meson behaves differently in different 
physical situations \cite{pennington} i.e. displaying different masses and 
lifetimes. This is also reflected by the wide mass range quoted by PDG: 
$400-1200$ MeV.
From the theoretical side, we can view the $\sigma$ meson as  
a Higgs particle in the context of the linear sigma model \cite{sigma} 
after the spontaneous
breaking of chiral symmetry, i.e., as the real part of this Higgs field whose 
vacuum expectation value breaks the chiral symmetry 
(the imaginary part resulting into pseudoscalars). Alternatively, 
one can look upon $\sigma$ as a low energy 
manifestation of the scale invariance breaking in the strong interaction 
\cite{scalebreaking} (the `identification' $\sigma \sim {\bf F}\cdot {\bf F}$ 
mentioned above is motivated by this scale invariance breaking 
since the latter is connected to the trace anomaly in QCD 
by which the dilaton current is not conserved; 
note again the analogy to the eta mesons where the connection to 
the gluon fields is through another anomaly, namely the $U(1)_A$). 
Hence the question: chiralon, dilaton or both?

In spite of these theoretical insights, there is no general consensus about 
the parameters of the $\sigma$ meson. The Nambu-Jona-Lasinio model 
\cite{NJL} would require a mass of roughly twice the constituent quark mass 
i.e. $600-700$ MeV. 
Weinberg's prediction in the framework of mended symmetry is $m_{\sigma}
\simeq m_{\rho}$ \cite{weinberg, svec}. The latter seems also to agree with 
values required to satisfy the Adler sum rule \cite{sumrule, svec}. 
Calculations using
Bethe-Salpeter approach \cite{burden1} give $m_{\sigma}\sim 750$ MeV 
which, however, is
rejected by the authors themselves \cite{burden2} to be the mass of the 
$\sigma$ meson on account of it being too heavy. 
Indeed, the pole values predicted in the unitarized chiral perturbation
theory \cite{oset} and the unitarized quark model \cite{toernqvist}, 
give the $\sigma$ mass around $400$ MeV, which is 
roughly in agreement with potential like models \cite{speth, beveren}. 
An even lower value for $m_{\sigma}$, 
namely $m_{\sigma}^{BW} \simeq 390^{+60}_{-36} $ MeV is 
reported by a very recent
observation of $J/\Psi \to \sigma \omega \to \pi \pi \omega$ \cite{wu}. 
Last but not least, we mention a mysterious low $\pi \pi$ mass 
enhancement at $310$ MeV which dates back to 1961 \cite{ABCold} 
and is since then coined ABC-effect. It has not vanished during the years, 
but continues to leave its fingerprints in several scattering 
processes \cite{ABCnew} till today.  

Above $1$ GeV, a resonance is often mentioned around 
$1200-1300$ MeV and was earlier known as 
$\varepsilon(1300)$ \cite{klempt, estabrooks}. It is often 
stressed that this is not the $f_0(1370)$ scalar because of different 
inelasticities
into $\pi \pi$ and $4\pi$. This problematic region borders to 
$f_0(400-1200)$
on the one hand and $f_0(1500)$ on the other. Therefore strong 
interference effects are to be expected. 
A strong support for a resonance around $1300$ is given in \cite{anisovich}. 
In \cite{ochs} the status of $f_0(1370)$ as a genuine resonance is doubted.

Putting everything together, we recognize three problematic regions in 
the $I(J^{PC})=0(0^{++})$ sector: $300-450$ MeV, $700-800$ MeV and 
$1200-1300$ MeV, where one finds strong hints/evidences for unusual 
activities/resonances. Actually these regions are often 
attributed to a single resonance, the $\sigma$. 
Reproducing the figure in \cite{pennington} with 
new data in Fig. 1, one sees that most data points are grouped around 
these three regions.

\begin{figure}[h]
\centerline{\vbox{
\psfig{file=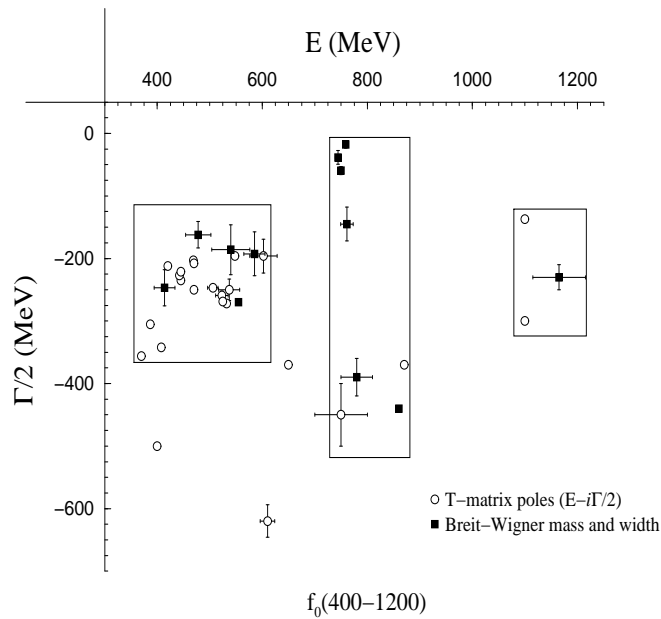,height=8cm,width=9cm}}}
\caption{Masses and widths of the scalar $\sigma$ meson as determined
by different groups as listed in PDG.}
\end{figure}

Having discussed the unflavoured mesons which we might find in the $\pi \pi$ 
system, we now turn our attention to strange resonances encountered 
in $\pi K$ scattering. We are not aware of any problems/controversies 
in the $I(J^{P})=1/2(1^-)$
sector and this  problem free zone is also confirmed nicely by our analysis.
There are some uncertain candidates in the  $I(J^{P})=1/2(2^+)$ 
spectrum around 
$2$ GeV \cite{pdg1}  and we add to this uncertainty two never-heard-of 
`candidates' around $1.7$ and beyond $2$ GeV.  We will discuss the 
reliability of these results in section 5. However, more importantly, it 
is again the scalar mesons $I(J^P)=1/2(0^+)$ 
which stir up controversies and it is again about the lightest strange scalar
which right now goes under the name $\kappa$. Whereas the first suggestions of 
$\sigma$ date back to early 60's \cite{earlysig}, one can trace back the 
$\kappa$ to early 70's \cite{earlyk}. As with the $\sigma$, the
$\kappa$ meson too has met with either an outright refutal for the 
basis of its existence \cite{pennington2} or support
together with determination of its mass and width. 
Two often quoted values for its mass are: $700-750$ MeV 
\cite{lighk} and close to $900$ MeV \cite{heavyk}. This resembles very much 
the situation of $\sigma$.  
\section{Time-delay and phase shift}
In this section we give the main formulae concerning time delay 
without derivation and refer the reader to literature for a more detailed 
account \cite{wigner,smith,books}. Intuitively, a {\it positive} 
time delay is a measure of 
how much a reaction is delayed, say due to an unstable intermediate state.
Close to the resonance region we would expect to find a {\it positive} 
peak in the energy distribution of time delay. 
Indeed, since in a resonant process we produce the
resonance on-shell, the process itself can be viewed in steps. The first is 
the resonance production at the space-time point $P_1=(t_1, x_1)$ followed 
by the
resonance decay at $P_2=(t_2, x_2)$ with $P_1 \neq P_2$ and $\Delta t =t_2-
t_1 > 0$ (non-localized processes can also be encountered in t-channel
\cite{tchannel} where one of the initial particles is a resonance).
{\it Negative} time delay occurs when the interaction is repulsive and/or a 
new decay channel opens up \cite{ng}. The measure for both has been found
by Wigner and Eisenbud \cite{wigner,wigner3} in the form of the 
energy derivative of the phase shift $\delta$ as,
\begin{equation}\label{1}
\Delta t = 2 \hbar {d\delta \over dE}\,
\end{equation}
which is valid for elastic processes \cite{wigner3}. Smith 
\cite{smith} generalized this to calculate the time delay
directly from the $S$ matrix (including also inelastic processes $ i \to j$)
giving, 
\begin{equation}\label{2}
\Delta t_{ij} = \Re e \biggl [ -i \hbar (S_{ij})^{-1} {dS_{ij} \over dE}
\biggr ] \, .
\end{equation}
With
\begin{equation}\label{3}
S_{kj} = \delta_{kj} \,+ \,2 \,i\, T_{kj} \,
\end{equation}
and
\begin{equation} \label{3a}
T_{kj} = \Re e T_{kj} \,+ \,i\,\Im m T_{kj}, 
\end{equation}
equation (\ref{2}) in the elastic case ($i=j$) can be recast into  
\begin{equation}\label{4}
S^*_{ii} \,S_{ii}\, \Delta t_{ii}\, =\, 2 \,\hbar\, 
\biggl[ \Re e \biggl({dT_{ii} \over dE}\biggr)\,+ \,2 \,\Re e T_{ii}\,\,
\Im m \biggl ({dT_{ii} \over dE}\biggr) \,-\, 2\, \Im m T_{ii}\,\, 
\Re e\biggl( {dT_{ii} \over dE}\biggr)\,
 \biggr],
\end{equation}
where obviously $\Delta t_{ii}$ is the same as in equation (\ref{1}). 
This can be checked by substituting $S = \eta e^{2 i \delta}$ (where 
$\delta$ is the real scattering phase shift and $\eta$ the inelasticity
parameter ($0 < \eta \leq 1$)) which gives, 
\begin{eqnarray}
\Delta t_{ii} &=& Re \biggl [ \,-i\hbar\, \biggl( \,2i
{d\delta \over dE} + {d\eta \over dE}\, {1\over \eta}\,
\biggr ) \biggr ] \nonumber \\
&=& \,2\hbar \, {d\delta \over dE}
\end{eqnarray}
Smith's 
formalism shows that Wigner's expression (\ref{1}) is valid for elastic 
reactions even if there are non-zero inelasticities. The generalization
of time delay with $N$-body interaction was done later in \cite{osborn}. The 
statistical connection to time delay mentioned in the introduction is given by
\cite{statistics}
\begin{equation} \label{5}
\sum_l \Delta n_l(E) =\sum_l {2l +1 \over \pi} {d\delta_l(E) \over dE}
\end{equation}
where $\Delta n_l$ is the difference in the density of 
states with and without interaction and $\delta_l$ is the phase 
shift in the $l$th partial wave.
Since $\Delta t$ can be positive as well as negative, some references 
call it collision time reserving the name time delay for the 
case $\Delta t >0$ and time advancement for $\Delta t < 0$.

It is well known that the lifetime of an unstable state depends in principle,
on its preparation, i.e. on the energy spread $\Delta E$
of the initial particles which produce the state. 
Most of the experiments producing hadron resonances certainly operate 
in the region where the inequality
\begin{equation} \label{6}
{\Delta E \over m}  < < {\Gamma \over m}
\end{equation}
with $\Gamma$, the width of the resonance, holds true.
In case of broad resonances, the Breit-Wigner distribution is not 
necessarily a good parametrization. Moreover, in such cases the mean 
lifetime is not $1/\Gamma$ \cite{joa, taylor}, but rather 
$\Delta t (E)$ (where $E$ is the energy available in the centre
of mass system), as given above \cite{newpapers}.
One can also put it in different words.
Since for a narrow resonance, the Breit-Wigner is the spectral 
function \cite{spectral,razka} which leads to the exponential decay law, 
it is reasonable to say that 
$\Delta t(E)$ is the right spectral function for a broad resonance (the Fourier 
transform of this spectral function gives the survival amplitude).
It is then clear that time delay is so to say tailored to study broad hadronic
resonances, especially the mesonic cases mentioned in the last section
(which we shall confirm in the next two sections).

Before applying the time delay method to the broad mesonic resonances, we
give here two words of caution. Since the collision time is given as 
{\it derivative} of the phase shift, a rather good quality of data
is required to avoid `false' bumps, i.e. signals in the time delay plots 
which are not genuine resonances but rather an artifact of the fit. 
Secondly, the
phase shift necessary to calculate the collision time, cannot be uniquely
determined through data. Indeed, one usually gets several solutions 
which then have to pass certain tests (involving crossing symmetry, 
unitarity, analyticity etc.) in order to decide which solution is the
physical one. Even then it is not so clear if one can avoid the so-called
continuum ambiguity \cite{contambiguity}. In our case, using time delay, 
a good check will always be if we can reproduce the well established
resonances.

In passing, we note that Wigner's intention in \cite{wigner} was to find
a criterion to distinguish between physical and unphysical phase shifts. 
He found
\begin{equation} \label{7}
\Delta t = 2 \hbar {d \delta \over d E} > -a
\end{equation}
which is a causality condition since a time advancement cannot be
arbitrarily large. In (\ref{7}) $a$ is interpreted as interaction range.

A legitimate question is about the relation between resonance parameters
defined as poles of the T-matrix and positive peaks in time delay. In
this connection we quote ref. \cite{peres} where the author says, 
``Moreover, if there is an appreciable time delay ($t''>t'$), the latter
should be interpretable as arising from the propagation of an unstable
intermediate particle. The above requirements, which are readily
generalized to multiple scattering processes, are sufficient to derive 
the pole structure of the S-matrix and the existence of antiparticles"
and ref. \cite{brans} where the authors state, ``However, the information
that gives an indication of a large time delay - typically the fast
traversal by the amplitude of a resonance circle as described in the 
sections ``Elastic Resonances" and ``Inelastic Resonances" - is the
same information whose extrapolation on to the unphysical sheet
informs us as to the existence of a pole". We can also perform a 
practical test by evaluating time delay using phase shifts calculated
within a model and comparing the time delay peaks with the pole values
of the T-matrix within that model. For instance, using the model of
Kaminski, Lesniak and Loiseau described in \cite{kaminmodel}, 
for the $\pi \pi$ elastic scattering phase shifts, we get the resonance
peaks shown in Fig. 2, which are in good agreement with the pole values
found within the model (see Table 3, of 
\cite{kaminmodel}).    
\begin{figure}[h]
\centerline{\vbox{
\psfig{file=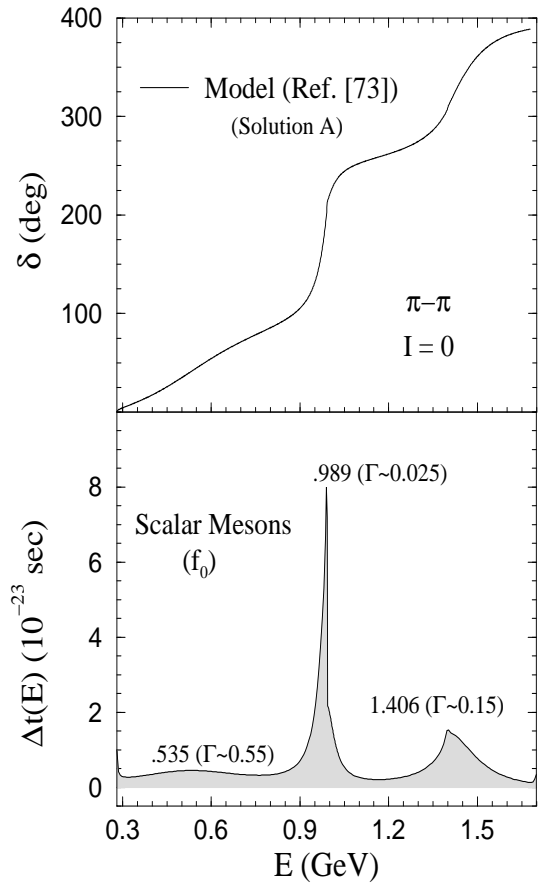,height=12cm,width=7cm}}}
\caption{Time delay plot of the scalar meson resonances evaluated using 
the s-wave phase shifts from the model calculation of \cite{kaminmodel} 
for $\pi \pi$ elastic scattering.}
\end{figure}

We would also like to draw the reader's attention to ref. \cite{lifshutz},
where a simple model of a coherent sum of seven resonances was made
to display the virtues of the time delay method. In the Argand diagram
plot, these resonances showed up misleadingly as one resonance whereas 
the time delay plot could differentiate between these resonances with
the peak values coinciding with the pole positions (see Fig. 1 and 
Fig. 3 in \cite{lifshutz}). This work indicates that the time delay
method is then well-suited for overlapping resonances. 

\section{Resonances in the $\pi \pi$ system}
We start our exploration of time delay and resonances, examining the
$\pi \pi $ phase shifts.
These phase shifts have been extracted with care and passed several
tests \cite{kaminski} (among others the Roy equations \cite{roy}).
Hence we have confidence that they represent the true physical situation 
(this is also justified because we do confirm all well-established resonances
in this case). A note about error bars is in order here. The errors in
the phase shifts get transformed into the errors in time delay and in
principle can shift the positions of the time delay peaks. However, 
the aim of the present work is not a precise determination of resonance
parameters but rather showing the usefulness of the time delay method. 
Hence we do not quote the errors in the peak positions, but we do make
a double check by using two different sets of phase shifts whenever
possible. 
\newline
\underline{$\rho$ mesons:}  This is an example of the analytical power of the
time delay method applied to resonance physics. As Fig. 3 nicely demonstrates, 
we find a prominent peak for the $\rho$ meson where one would expect it to be, 
at $765$ MeV. The phase shifts are from \cite{rhodata}. 
\begin{figure}[h]
\centerline{\vbox{
\psfig{file=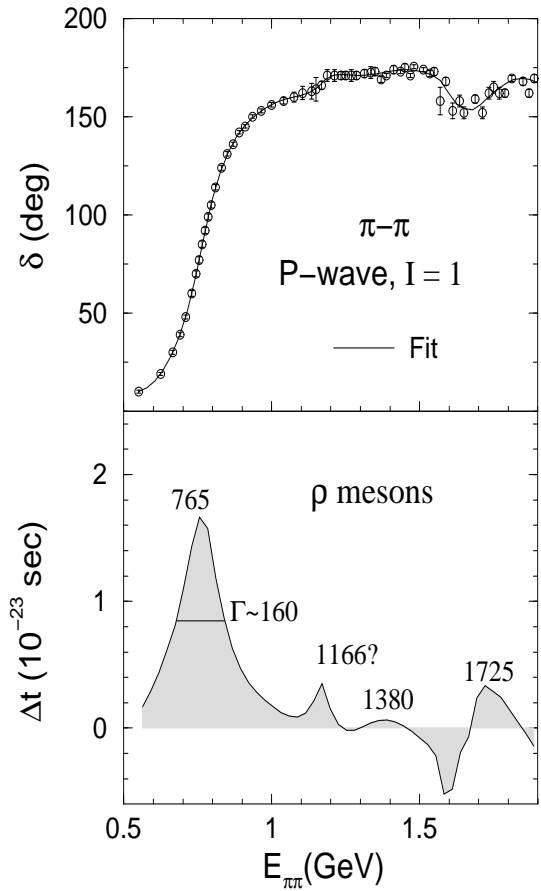,height=12cm,width=7cm}}}
\caption{Time delay plot of the $\rho$ resonances evaluated from a 
fit to the p-wave phase shifts in $\pi \pi$ elastic scattering.}
\end{figure}
We identify the peak at $1380$ MeV with $\rho (1450)$. The 
PDG's estimate (`educated guess') for this resonance is  \cite{pdg1}
$1465$ MeV. However, in the $\pi \pi$ mode the mass range listed is $1292-
1406$ MeV with an average value of $1370$ MeV. A similar analysis can be 
applied to the next peak at $1725$ MeV which is due to $\rho(1700)$. 
The peak value is close to the PDG estimate of $1720 \pm 20$ MeV.
According to PDG, the mass for this resonance in the $\pi \pi$ mode
ranges between $1590$ and $1838$ MeV. Having confirmed the three known mesons
in the time delay plot, it was worrying to find an additional peak at
$1166$ MeV. It does not seem that this is an artifact of the fit or due to
poor quality of the data and we had to accept this signal as a genuine one.
We found independent support in \cite{bartalucci, achasov, pick, donnachie1}
resulting in an accumulated evidence for a $\rho$ resonance around $1200$ MeV.
Note that the sources of this evidence are quite different.  
\newline
\underline{Scalar ($f_0$) mesons:}
\begin{figure}[h]
\centerline{\vbox{
\psfig{file=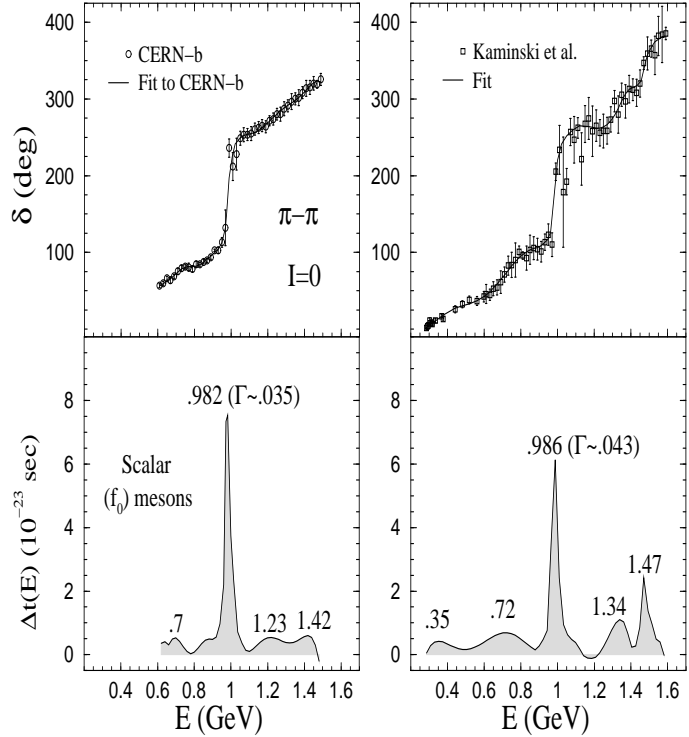,height=10cm,width=9cm}}}
\caption{Scalar meson resonances as seen in the time delay evaluated
from fits to s-wave phase shifts in isoscalar $\pi \pi$ scattering.}
\end{figure}
The phase shifts in Fig. 4 are from \cite{kaminskidata} and \cite{cernB}. The
time delay calculated from the two different sets of phase shifts 
gives a similar qualitative
picture which shows that the underlying physics should be taken seriously.
The peak on the far right $1.42 (1.47)$ GeV is  a signal of $f_0(1500)$
whose possible mass values listed in PDG are between $1.4$ and $1.6$ GeV.
This resonance is overlapping from the left with $f_0(1370)$ for which we find
peaks at $1.23 (1.34)$ GeV (PDG quotes $1.2-1.5$ for the pole position and 
also for the Breit-Wigner value). The sharp peaks at $0.982 (0.986)$ are
of course the well established $f_0(980)$ bordering at a resonance structure 
at $0.7$ GeV, which in turn in one of the cases overlaps with a bump at $0.35$ 
GeV. The individual peaks are not well separated due to the large
widths of scalar resonances. Nevertheless, the time delay method is able
to distinguish between these overlapping cases. 
Due to the successful separation of standard 
cases, we do not think that the 2 peaks below $1$ GeV (350 and 700 MeV)
are accidental or due to the fit and/or data.
These two regions at $400$ and $700$ MeV are also the most quoted
in connection with the $\sigma$ meson mass. It is worth stressing that we have
recovered these regions here in one and the same reaction.

\begin{figure}[h]
\centerline{\vbox{
\psfig{file=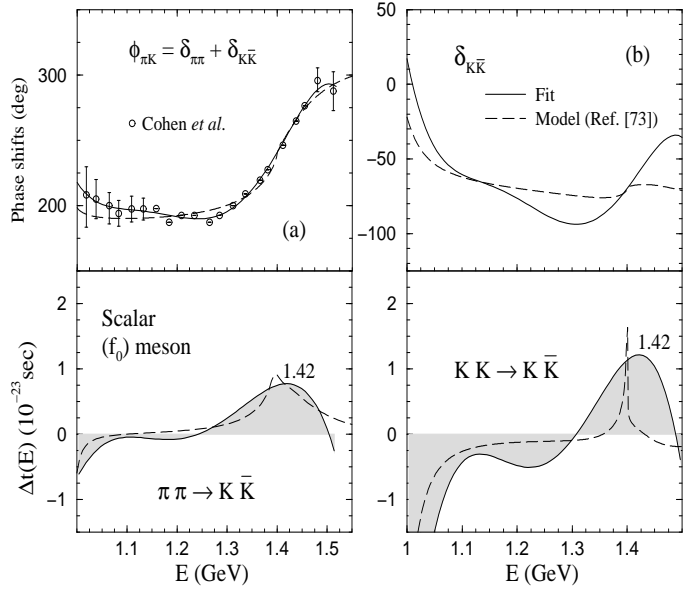,height=8cm,width=9cm}}}
\caption{Energy dependence of the s-wave phase shifts and time delay 
in the (a) $\pi \pi \rightarrow K \bar{K}$ and (b)  
$K \bar{K} \rightarrow K \bar{K}$ reactions. The phase shifts  
$\delta_{K \bar{K}}$ (solid lines in (b)) have been determined by 
subtracting the fit to the 
CERN-b $\pi \pi$ phase shift data in Fig. 4 from $\phi_{\pi K}$ in the
above figure (a). Dashed lines show the same quantities in the model 
calculation of ref. \cite{kaminmodel}.}
\end{figure}
At the end of this section, it is instructive to subject the time delay
method to yet another test. The resonances found in the elastic 
scattering process, $\pi \pi \rightarrow \pi \pi$, should also manifest
themselves in the coupled channel reactions, $\pi \pi \rightarrow 
K \bar{K}$ and $K \bar{K} \rightarrow K \bar{K}$, provided they have
an appreciable branching ratio for decaying into $K \bar{K}$. The plots
in Fig. 5 are just the results of such a test in the mass range from 
1 to 1.5 GeV. In Fig. 5a, we show the data on $\phi_{\pi K}$ ($\phi_{\pi K}= 
\delta_{\pi \pi} + \delta_{K \bar{K}}$, where $\delta_{\pi \pi}$ and
$\delta_{K \bar{K}}$ are the elastic scattering phase shifts in the
$\pi \pi$ and $K \bar{K}$ channels) obtained in \cite{wicklund, cohen}. 
We make a fit to this data and subtract from it, the fit made earlier 
in Fig. 4 to the CERN-b $\pi \pi$ data, to obtain the $K \bar{K}$ 
phase shift (solid line) shown in Fig. 5b. The dashed lines in this 
figure are the phase shifts obtained in a model calculation in 
\cite{kaminmodel}. The resonance found at 1.42 GeV (peaks in
shaded regions) using $\phi_{\pi K}$ and $\delta_{K \bar{K}}$ is to
be associated with $f_0$(1370). We can see that by using the model
phase shifts of \cite{kaminmodel} which give a peak at 1.4 GeV. Note
that the data sets of $\pi \pi \rightarrow \pi \pi$ and 
$\pi \pi \rightarrow K \bar{K}$ reactions are different and we do not 
expect the mass values from the time delay peaks in 
$\pi \pi \rightarrow \pi \pi$ (Fig. 4) and 
$K \bar{K} \rightarrow K \bar{K}$ (Fig. 5) to be exactly the same. 
Note also that nothing can be inferred about $f_0$(980) and $f_0$(1500) 
since both are just at the edge of the data set in Fig. 5.

The data in Fig. 5 has the advantage that one can compare the result 
from time delay with the speed plot shown in \cite{wicklund,cohen}. The
speed plot is actually the energy distribution, $SP(E)$, 
defined as,
\begin{equation}
SP(E) = \biggl| {dT \over dE} \biggr |
\end{equation}
where $T$ is the energy dependent complex $T$-matrix. In 
\cite{wicklund,cohen}, $SP(E)$ has been calculated (Fig. 2 in 
\cite{wicklund} and Fig. 28(b) in \cite{cohen}) and the speed plot
peak appears at 1.425 GeV, which is in excellent agreement with
our finding of the time delay peak in Fig. 5.      
\section{Resonances in the $\pi K$ system}
We now perform a similar analysis as above in the strange meson sector. 
The results confirm all established mesons, the $\kappa$ meson and 
hint in addition to the existence of new strange mesons.
\newline
\underline{Strange ($K_0^*$) scalar mesons:}
The only well established strange scalar is $K^*_0(1430)$. However, over the 
years there have been claims about the existence of a lighter strange scalar
called $\kappa$ \cite{earlyk, lighk, heavyk}. 
\begin{figure}[h]
\centerline{\vbox{
\psfig{file=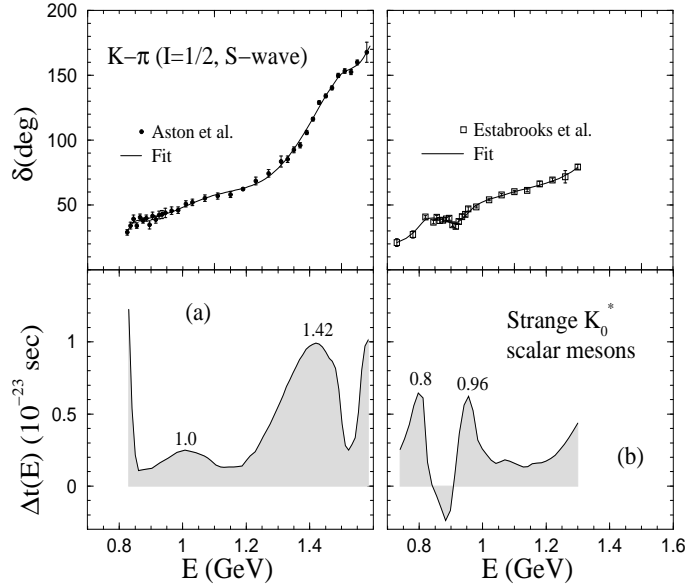,height=8cm,width=9cm}}}
\caption{Time delay plots evaluated from the phase shifts in 
s-wave, isospin 1/2 $K \pi$ scattering, displaying the strange scalar
mesons.}
\end{figure}
Its mass is thought to be within the range $660-900$ MeV \cite{lighk, heavyk}.
The situation resembles the case of the light unflavoured mesons, not only
regarding its existence, but also its wide mass range thought to be possible. 
It should then come as no surprise, when the broad mass range 
gets resolved into several peaks. Two sets of phase shifts from 
\cite{strangescalaraston} and \cite{strangescalarestabrooks} 
are available to study these scalar mesons. Though both of them cover 
only a small energy region, we can see that they are complementary 
in a sense. In Fig. 6a we see a clear
signal for $K^*_0(1430)$, whereas in Fig. 6b, 
the tendency for a peak in this mass range is present. The peak at $1$ GeV 
(Fig. 6a) is also accompanied by a tendency towards a lower peak. 
This is confirmed by the other time delay plot in Fig. 6b where we see two 
peaks at $0.8$ GeV and $0.96$ GeV. 
These two peaks can be interpreted as a signal of the $\kappa$ meson 
in time delay plots. Since the two peaks are very close to each other, 
they could have been confused as one single resonance.
\newline
\underline{Strange ($K^*$) vector mesons:}
The strange vector mesons have been so far a no-problem zone for
resonance spectroscopy. The time delay plots in Fig. 7 calculated from the 
phase shift given in \cite{strangescalaraston} do confirm this. Clear signals
for $K^*(892)$, $K^*(1410)$ and $K^*(1680)$, with PDG mass estimates
of $m=892$, $1414$ and $1717$ MeV respectively, are seen. 
\begin{figure}[h]
\centerline{\vbox{
\psfig{file=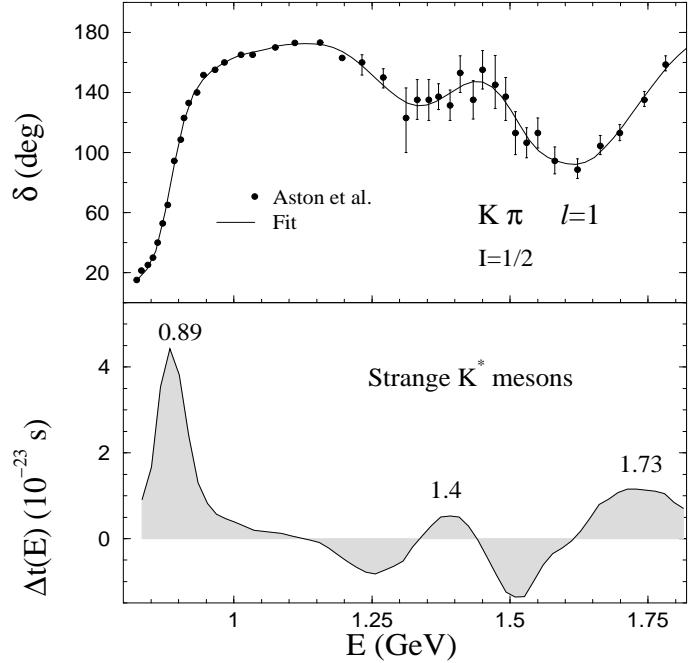,height=9cm,width=9cm}}}
\caption{The strange $K^*$ mesons as observed in time delay evaluated from
a fit to the p-wave phase shifts in the isospin 1/2, $K \pi$ scattering.} 
\end{figure}
\newline
\underline{Strange ($K_2^*$) tensor mesons:}
 Alone a visual inspection of the two phase shift solutions  from
\cite{Kl2data} reveals that
this case is different from the previously discussed cases. The visual
inspection also reveals that there will be four peaks in the corresponding time 
delay plots which is indeed confirmed in Fig. 8. Note that the two solutions
give us within errors the same spectrum of the strange spin-two 
resonances. The first peaks in Figs 8a and 8b can be clearly attributed to
$K^*_2(1430)$. The peaks at $2.0$ and $2.18$ GeV are a confirmation of
$K^*_2(1980)$ which is listed in PDG, but omitted from its Summary
Table (see also \cite{aston2}). We find additional peaks around $1.73$ and
$2.4$ GeV. In \cite{Kl2data} the authors note that ``... , but above $1$
GeV, there is suggestive evidence for phase motion, though little structure 
in the magnitude.'' We believe that we have now quantified the expression 
`phase motion' through our time delay analysis. The additional peaks which 
we found cannot be easily waived away once we accept the phase shifts where
the quality of the data is sufficiently solid to support them. In the 
$K \pi$ system, one sees more regions of negative time delay. This happens
very often in the presence of increased inelasticities. However, it was
also pointed out by Wigner \cite{wigner} that negative time delay is
unavoidable over the whole energy region.       
\begin{figure}[h]
\centerline{\vbox{
\psfig{file=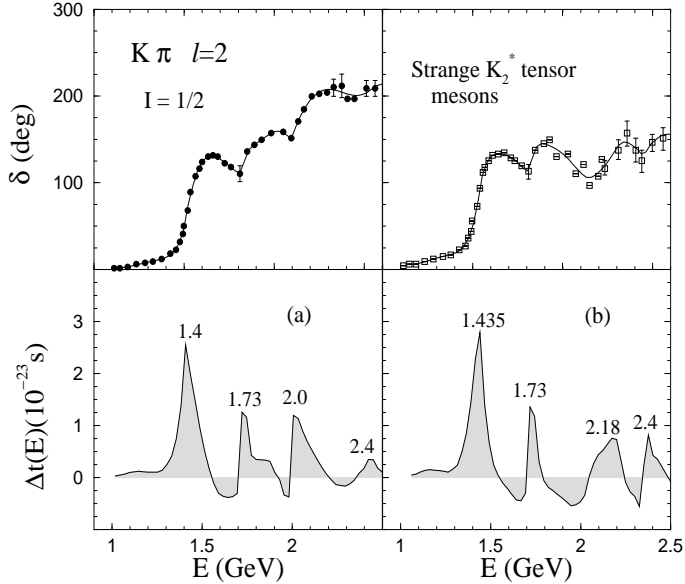,height=8cm,width=9cm}}}
\caption{The strange $K_2^*$ tensor mesons as observed in time delay 
evaluated from a fit to the d-wave phase shift solutions of 
ref. \cite{Kl2data} for the isospin 1/2, $K \pi$ scattering.} 
\end{figure}
\section{Conclusions}
In the present work, we have used the method of time delay to investigate
the resonances occurring in $\pi \pi$ and $\pi K$ reactions. 
Time delay (in connection with hadronic resonances) has been advocated 
since fifty years, but to our knowledge never used in practice. 
Given the available phase shifts, the outcome of such
an analysis is beyond our control, and in principle, it was possible 
that our conclusions regarding the usefulness of time delay in resonance 
physics would come out to be negative. However, the contrary is the case. 
In our time delay plots, 
the well established resonances always appeared as prominent peaks, 
in all the cases we examined, which we believe is already quite remarkable.
Apart from this we also find additional information. In the scalar sector, 
the signals at $350$ and $720$ MeV are in regions
claimed also by other authors/groups to be regions of resonance activity.
In the strange scalar sector we confirm the $\kappa$ meson; however, we get two
neighbouring peaks. Since these peaks are very close to each other, 
we cannot say with certainty if they are really two resonances or artifacts of 
the fit. In the $\rho$ mesonic sector, we find evidence of a resonance
around $1200$ MeV which again has been and is a matter of debate (though
it has been omitted by PDG in 2002 for lack of evidence, in the same year 
several groups found the evidence for its existence using different methods). 
The $K^*$ ($J=1$) resonances would be the case par excellence,  
since no problems are known in this sector and all that we found are the well
established cases too. The only sector where we found never-heard-of 
resonances (again in addition to known cases) is the $K^*_2$ group. 
The phase shifts in this particular case look distinctly different from the
other cases we have examined. Taking the given phase shifts
at face value, it would certainly be worthwhile looking into the possibility
of two new strange tensor resonances. 
This is supported by the fact that two different
phase shift solutions yield the same resonances in the time delay plot.

In general, it seems to us that the time delay method is a useful 
analytical tool in studying resonances. It certainly is not an exclusive 
tool, but taken together with other methods it allows us to have 
additional insight into the complicated matter of overlapping 
and/or broad resonances, as it can apparently resolve individual peaks. 
For instance, in the enigmatic scalar sector, two phase shifts from 
different groups (not two different solutions) show the same behaviour:
three peaks around $400$, $700$ and $1200$ MeV, rather than one broad
resonance structure. 
\vskip0.5cm                                                
{\elevenbf \noindent Acknowledgments} \newline
We wish to thank V. Krey and J. C. Sanabria for useful
information and discussions. We are very thankful to 
Prof. R. Kaminski for providing the programs and phase shifts 
generated in their model \cite{kaminmodel}.
                                                         
\end{document}